\documentclass[sigconf]{acmart}

\AtBeginDocument{%
  }

\setcopyright{acmlicensed}
\copyrightyear{2018}
\acmYear{2018}
\acmDOI{XXXXXXX.XXXXXXX}

\acmConference[X]{Make sure to enter the correct conference title from your rights confirmation emai}{X}{X}

\usepackage{multirow}
\usepackage{enumitem}
\usepackage{titlesec}
\titlespacing*{\section}{0pt}{2pt}{2pt}
\titlespacing*{\subsection}{0pt}{2pt}{2pt}
\usepackage{bbding}
\usepackage{subfigure}
\usepackage{caption}
\usepackage{fancyhdr}
\usepackage{amsmath}
\usepackage{graphicx}
\usepackage{fancyhdr}
\usepackage{lipsum} 
\usepackage{threeparttable}

\begin{document}

\title{RRAM-Based Bio-Inspired Circuits for Mobile Epileptic Correlation Extraction and Seizure Prediction}


\author{Hao Wang$^1$, Lingfeng Zhang$^1$, Erjia Xiao$^1$, Xin Wang$^1$, Zhongrui Wang$^2$\textsuperscript{\textdagger}, Renjing Xu$^1$\textsuperscript{\textdagger}}
 \affiliation{
   \institution{The Hong Kong University of Science and Technology(Guangzhou)$^1$, The University of Hong Kong$^2$}
   \country{renjingxu@hkust-gz.edu.cn, zrwang@eee.hku.hk}
 }

 








\settopmatter{printacmref=false} 
\renewcommand\footnotetextcopyrightpermission[1]{}
\renewcommand{\shortauthors}{Hao Wang, et al.}

\begin{abstract}

Non-invasive mobile electroencephalography (EEG) acquisition systems have been utilized for long-term monitoring of seizures, yet they suffer from limited battery life. Resistive random access memory (RRAM) is widely used in computing-in-memory(CIM) systems, which offers an ideal platform for reducing the computational energy consumption of seizure prediction algorithms, potentially solving the endurance issues of mobile EEG systems. To address this challenge, inspired by neuronal mechanisms, we propose a RRAM-based bio-inspired circuit system for correlation feature extraction and seizure prediction. This system achieves a high average sensitivity of 91.2\% and a low false positive rate per hour (FPR/h) of 0.11 on the CHB-MIT seizure dataset. The chip under simulation demonstrates an area of approximately 0.83 \text{mm}$^2$ and a latency of 62.2 \textmu s. Power consumption is recorded at 24.4 mW during the feature extraction phase and 19.01 mW in the seizure prediction phase, with a cumulative energy consumption of 1.515 \textmu J for a 3-second window data processing, predicting 29.2 minutes ahead. This method exhibits an 81.3\% reduction in computational energy relative to the most efficient existing seizure prediction approaches, establishing a new benchmark for energy efficiency.

\end{abstract}



\maketitle


\section{Introduction}

Seizure disorders are widely predicted and analyzed using multichannel electroencephalography (EEG). However, traditional clinical EEG acquisition relies on large-scale hardware systems, which have limited portability and affordability\cite{biondi2022noninvasive}. To address this challenge, new noninvasive mobile EEG acquisition systems have been developed over the last two decades. However, most mobile EEG systems with built-in wireless/bluetooth data transmission modules and seizure detection algorithms fall short in terms of battery life\cite{sokolov2020tablet}.

Traditional seizure detection and prediction algorithms require a complex feature extraction process but support lightweight learning models such as Support Vector Machines (SVM), k-Nearest Neighbors (kNN), and Random Forest (RF) classifiers\cite{siddiqui2020review}. Recently, deep learning methods have been employed for seizure prediction, eliminating the need for complex feature extraction processes. Instead, the network directly undertakes feature analysis and seizure predicting. However, these methods consume significant computational resources\cite{ibrahim2022deep}.

RRAM has been leveraged to enhance computing efficiency through its in-memory computing capabilities\cite{li2022seizure}. Although RRAM-based seizure prediction has successfully reduced the computational energy consumption of deep learning methods, it has not yet been explored for traditional approaches, which combine physical feature extraction with simple models. Correlation is one of the commonly used EEG features\cite{williamson2012seizure}. Phase-change memory (PCM)\cite{sebastian2017temporal} and second-order memristors\cite{wu2021neural} have been utilized in the study of data correlations.

To minimize the computational energy consumption of seizure detection algorithms, inspired by neuronal correlations, we build a RRAM-Based circuit system for the extraction of EEG correlation features, along with an on-chip neural network system to predict the seizure. Fig.~\ref{fig: compare} illustrates a comparative analysis between our work and related studies. Our method significantly reduces the computational complexity of the EEG prediction algorithm while ensuring high predictive performance. Our specific contributions are as follows:

\begin{figure}[!t]
\centering
\captionsetup{font= small} %
\includegraphics[width=1\columnwidth]{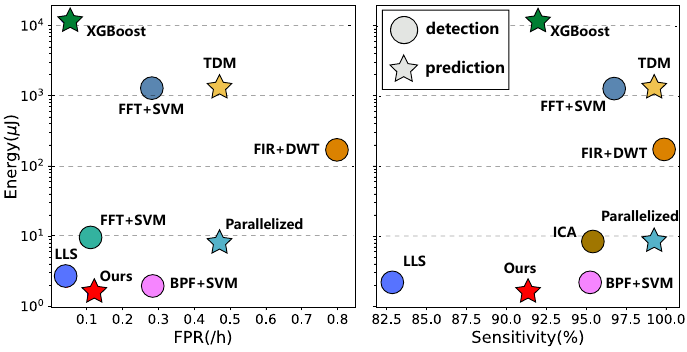}
\caption{Comparison diagram.}
\label{fig: compare}
\end{figure}
\vspace{-2pt}
\begin{itemize}[leftmargin=*]
  \item \textbf{RRAM-Based Correlation Extraction:} We initially proposed RRAM-based circuits for extracting EEG correlation features and successfully utilized them for seizure prediction.
  
  \item \textbf{On-chip Energy-Efficient Algorithm:} We have constructed a system for feature extraction and a two-layer epileptic seizure prediction artificial neural network (ANN) using only two 1T1R arrays, sized 18×18 and 18×2. According to our circuits operational logic, this system achieves the lowest energy consumption(1.515 \textmu J) for current seizure prediction tasks on the CHB-MIT epilepsy dataset, while maintaining high predictive performance.
  
  \item \textbf{Operational Logic for Specialized Application:} We have developed specialized system operational logic for seizure prediction, which includes unique EEG signal encoding methods, chip architecture reuse strategies, and DAC-less ANN network hardware mapping rules.
\end{itemize}
\vspace{-1pt}
\begin{figure}[!t]
\centering
\captionsetup{font= small} %
\includegraphics[width=\columnwidth]{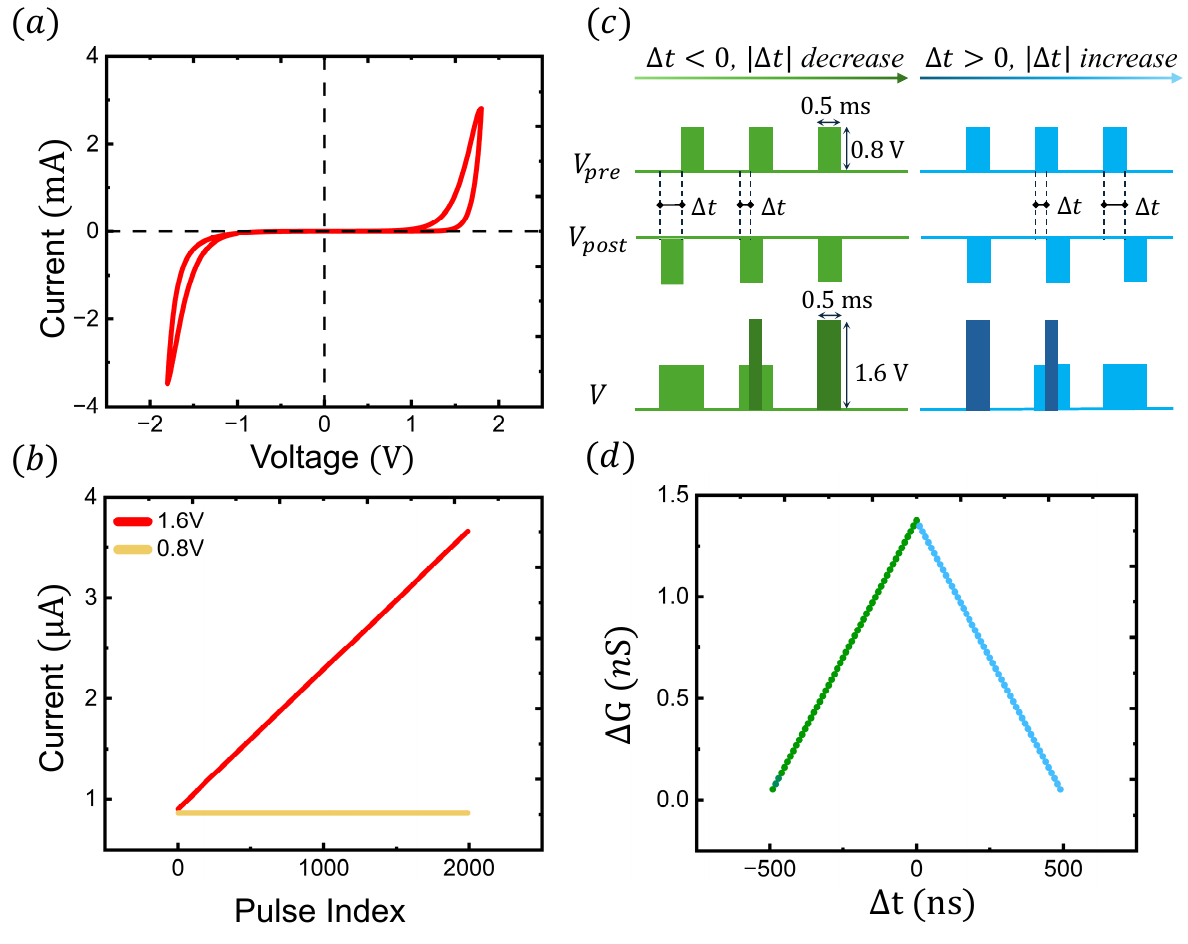}
\caption{RRAM Model Simulation Diagram. (a) Graph of the current through the memristor versus the voltage difference across its terminals. (b) Relationship between the current through the memristor and the number of pulses, upon inputting 2000 pulses of 1.6V, 500ns, and 0.8V, 500ns. (c) Diagram of the simulation logic used for (d). (d) Graph showing the relationship between pulse overlap ($\lvert \Delta t \rvert$) and the rate of change in conductance ($\Delta G$).}
\label{fig: simulation model}
\end{figure}
\vspace{-5pt}
\section{Background and Related Works}
This section begins with an overview of studies on epilepsy algorithms. Then, the discussion turns to the application of memristors in analyzing epilepsy data.

\begin{figure*}[!t]
\centering
\captionsetup{font=small} %
\includegraphics[width=\textwidth]{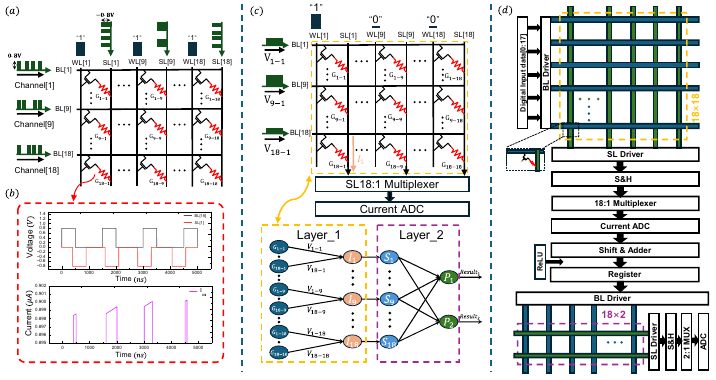}
\caption{EEG Circuit and System Logic. (a) EEG-Extracting Circuit and Logic. (b) Simulation partial results of the EEG-Extracting phase using RRAM ($G_{18-1}$) in LTspice. (c) Schematic of the ANN and circuit logic of the first layer network. (d) On-chip EEG-Computing system}
\label{fig: chip circuit architecture}
\vspace{-5pt}
\end{figure*}
\subsection{Seizure Algorithms}
Traditional seizure prediction algorithms typically necessitate intricate feature extraction from EEG signals. Techniques commonly utilized encompass correlation analysis \cite{williamson2012seizure}, wavelet transform \cite{liu2012automatic}, fast Fourier transform (FFT) \cite{chen2014automatic}, and least squares parameter estimators \cite{chisci2010real}. Following this feature extraction, simple learning models, such as support vector machines (SVM), are employed to predict seizures based on the extracted features. In traditional algorithms, the feature extraction process represents the most computationally demanding component.

Recently, the deep learning algorithms, specifically Convolutional Neural Networks (CNN) and Recurrent Neural Networks (RNN), has garnered significant attention in seizure prediction tasks due to their ability to perform feature extraction without complex algorithms. Notable contributions include Hisham Daoud et al.\cite{daoud2019efficient}, who employed CNNs for EEG feature extraction and RNNs for seizure prediction, achieving a high sensitivity of 99.6\% and a low FPR/h of 0.004 on the CHB-MIT dataset, with predictions made up to one hour in advance. Ranjan Jana et al.\cite{jana2021deep} utilized CNNs for both feature extraction and prediction on the CHB-MIT dataset, optimizing the number of channels from 22 to 6, resulting in a sensitivity of 97.83\% and 0.0764 FPR/h, albeit with a prediction window of only 10 minutes.

\subsection{Memristors and Epilepsy Analysis}

Memristors have been explored in the study of epilepsy-related research. Zhengwu Liu et al.\cite{liu2020neural} utilized memristor arrays for filtering and recognizing epileptic signals on the Bonn University dataset, achieving a high recognition accuracy of 93.46\%. Corey Lammie et al.\cite{lammie2021towards} were the first to implement a deep learning system for seizure prediction on memristors, achieving a prediction sensitivity of 77.4\% on the CHB-MIT dataset. Chenqi Li et al.\cite{li2022seizure} employed a low-latency parallel CNN architecture to perform prediction tasks on the CHB-MIT and SWEC-ETHZ epilepsy datasets, achieving seizure detection sensitivities of 99.24\% and 98.22\%, but with FPR/h as high as 0.47 and 0.99.
Notably, Abu Sebastian et al.\cite{sebastian2017temporal} used PCM for studying the correlation of weather information and image data. Yuting Wu et al.\cite{wu2021neural} employed Second-Order Memristors for researching neuron correlations. Currently, there has been no work that employs traditional non-volatile RRAM for data correlation analysis and uses it for seizure prediction tasks.

\section{System Implementation}

This section begins with the principles of RRAM, discussing RRAM-based overlap correlation, and outlines the construction and operational architecture of a chip system for processing 18-channel epilepsy EEG data. The EEG-Extracting section introduces the use of memristors for extracting correlation features, while the EEG-Computing section details the neural network structure for seizure prediction and the logic for on-chip operation.

\subsection{Memristor Model}

In this work, we utilize RRAM devices to extract spike overlap correlations. We will use the WOx-based memristor model proposed in\cite{cai2019fully} to construct our model:
\begin{equation}
I = (1 - w)\alpha[1 - \exp(-\beta V)] + w\gamma \sinh(\delta V)
\label{eq:equation1}
\end{equation}
\begin{equation}
\frac{dw}{dt} = \lambda \sinh(\eta V)
\label{eq:equation2}
\end{equation}

\noindent Parameters $\alpha$, $\beta$, $\gamma$, $\delta$, $\lambda$, $\eta$ are all positive-valued parameters determined by material properties. $\tau$ is the diffusion time constant. Additionally, $0 \leq w \leq 1$, $\alpha = 9 \times 10^{-7}$, $\beta = 4$, $\gamma = 2.8 \times 10^{-7}$, $\delta = 6$, $\lambda = 0.045$, $\eta = 6$.

The team provides a SPICE simulation model\cite{chang2011synaptic} for the relevant framework, which has been widely used for RRAM simulations. Using the LTspice software, a simulation system was constructed, yielding the imagery depicted in Fig.~\ref{fig: simulation model}. Part (a) illustrates the quasi-static I-V, with the memristor threshold approximately at 1V. In part (b), linear memristor programming was conducted using 2000 pulses of 1.6V, 500ns and 0.8V, 500ns, respectively; the graph indicates that the 0.8V pulses nearly do not instigate any change in the memristor. In part (c), $V_{pre}$ and $V_{post}$ are respectively applied to the two ends of the memristor, and the differential voltage V is $V_{pre} - V_{post}$. In the diagram, as the $ \Delta t $ changes, the overlap of the pulses first increases and then decreases. In Part (d), stimuli are applied to the memristor following the protocol established in Part (c). It is observed that the change in memristor conductance ($\Delta G$) is positively correlated with the increase in the area of pulse overlap ($\lvert \Delta t \rvert$). This indicates that within a 500 ns timeframe, a greater overlap area of the pulses signifies a higher correlation between the data from the two channels during this period.

\subsection{RRAM-Based overlap correlation}

The Pearson Correlation Coefficient ($PCC$)~\cite{cohen2009pearson}, a widely utilized indicator of linear correlation, quantifies the extent to which two variables are interrelated linearly. It gauges how variations in one variable align with changes in another. This relationship is mathematically expressed as:
\begin{equation}
PCC(x_1, x_2) = \frac{\sum_{n=1}^N (x_{1,n} - \overline{x_1})(x_{2,n} - \overline{x_2})}{\sqrt{\sum_{n=1}^N (x_{1,n} - \overline{x_1})^2 \sum_{n=1}^N (x_{2,n} - \overline{x_2})^2}}
\label{eq:equation3}
\end{equation}

Specifically, $N$ is the number of samples. $x_1$ and $x_2$ represent the series being analyzed, with $\overline{x_1}$ and $\overline{x_2}$ being their respective means. $PCC$ can vary from $-1$, indicating a perfect negative linear relationship, to $+1$, signifying a perfect positive linear relationship.
Synaptic connection strength can be estimated by evaluating the effect of the discharge of a given neuron on the spike probability of another neuron\cite{wu2021neural}. Inspired by biological relevance, we propose an RRAM-Based Overlap Correlation. Within a 500 ns time window, the memristor emulates a biological synapse where a greater overlap in the pulse signals at both ends of the memristor signifies a larger conductance change ($\Delta G$), indicating stronger signal correlation. Over time, the conductance change($\Delta G_k$) will accumulate, indicating that the conductance value measured at the end of a period can represent the correlation between the continuous pulse signals at the terminals of the memristor during this interval. This can be represented by the following formula:
\vspace{-2pt}
\begin{equation}
G_t = \sum_{k=0}^{t} \Delta G_k + G_0
\label{eq:equation4}
\end{equation}
Here, $G_t$ represents the cumulative value at time $t$, $\Delta G_k$ are incremental changes over time, and $G_0$ is the initial value. A larger measured conductance value at time $t$ indicates a stronger signal correlation between the two ends over the period from $0$ to $t$.
\begin{figure*}[!t]
  \centering
  \captionsetup{font=small} %
  \includegraphics[width=0.9\textwidth]{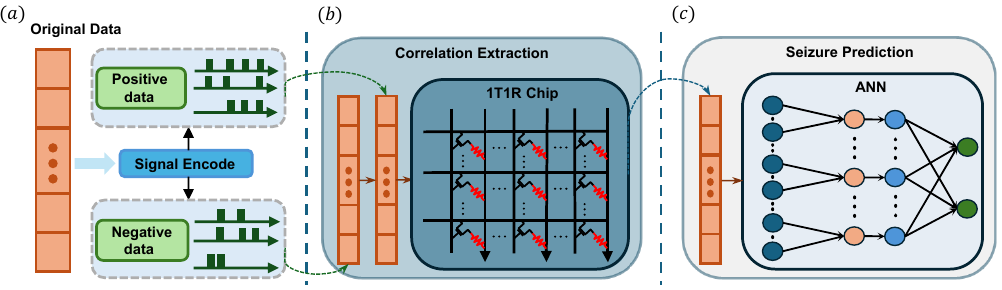}
  \caption{RRAM-Based System Framework. (a) EEG Signal encoding. (b) Correlation Extraction. (c) Seizure Prediction}
  \label{fig: framework}
\end{figure*}
\vspace{-2pt}
\subsection{EEG-Extracting}
Our experiment utilizes the CHB-MIT dataset\cite{shoeb2009application}, which contains epileptic EEG data from 18 channels. Consequently, the correlation features comprise 18×18 data. Capitalizing on the overlapping correlation properties of RRAM, we proposed a chip design dedicated to extracting EEG correlation features and predicting seizure, as demonstrated in Fig.~\ref{fig: chip circuit architecture}.

Fig.~\ref{fig: chip circuit architecture} (a) illustrates the detailed operational logic of an 18×18 array during the EEG-Extracting phase. All WLs are set to `1', fully activating the transistors, thereby enabling the one-to-one input of EEG data from 18 channels along rows and columns at the BLs and SLs sides, respectively. The BLs are fed with positive pulses of 0.8V, 500ns, while the SLs side receives negative pulses of -0.8V, 500ns. Each memristor sequentially records the cumulative overlap correlation values between pairs of EEG channels. Ultimately, the correlation magnitudes of the 18 EEG channels are mapped to the sizes of the 18×18 conductance values within the array.

Based on the correspondence of channel overlap correlation, we have assigned designations to the memristors, ranging from $G_{1-1}$ to $G_{18-18}$, totaling 324 identifiers. We conducted a preliminary test of the system using the LTspice software. Fig.~\ref{fig: chip circuit architecture} (b) presents a partial simulation image of the memristor labeled $G_{18-1}$. From this, we observe that the change of the memristor conductance value varies with the size of the pulse overlap area, corroborating the characteristics shown in Fig.~\ref{fig: simulation model} (b).
\vspace{-2pt}
\subsection{EEG-Computing}
Following the EEG-Extracting phase, in pursuit of minimal energy consumption, we explored the use of a two-layer neural network specifically for the task of seizure prediction. The first layer of the network, which is directly implemented on the EEG-Extracting circuits. The second layer is mapped onto the RRAM chip with specific rules.

Fig.~\ref{fig: chip circuit architecture} (c) presents the neural network model employed for the seizure prediction, and succinctly describes the implementation of the first network layer using the same chip as that utilized for EEG-extracting. Following the EEG-Extracting phase, the memristor conductance values-representing EEG correlations are utilized 
as inputs for the ANN. Considering the voltage threshold of 1V for the memristor, the weights of the first layer of the neural network are trained within a range of 0 to 1. Since the EEG correlations are already stored on-chip, it is only necessary to convert the first layer network weight values into corresponding voltage values and input them sequentially from the BLs side. Subsequently, by employing the multiplexer on the SLs side and the ADC module, the accumulated currents can be read out column by column, completing the computation of the first neural network layer. 

Based on the circuits provided by the classic memristor chip\cite{yao2020fully}, we introduce an EEG-Computing circuit for seizure prediction, as illustrated in Fig.~\ref{fig: chip circuit architecture} (d). To achieve robust computation, the input weights of the first layer network are supplied by digital signals. The output of this network layer is obtained through the logic operation depicted in Fig.~\ref{fig: chip circuit architecture} (c). Due to the use of digital signals for computations, a Shift\&Adder structure is required after the ADC module. A ReLU activation function is employed between the two network layers. The second network layer utilizes standard on-chip RRAM In-Memory Computing execution rules\cite{zhang2023edge}. Mapping the 0-1V voltage weights of the first layer to digital signals can reduce the utilization of Digital-to-Analog Converters (DACs), thereby decreasing chip area and lowering energy consumption. However, this introduces the use of Shift\&Adders and incurs increased latency.

\begin{figure*}[!t]
\centering
\captionsetup{font=small} %
\includegraphics[width=\textwidth]{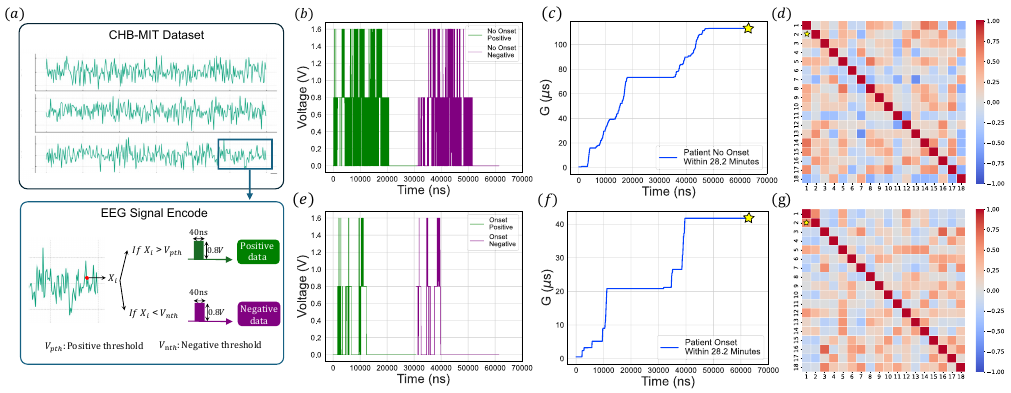}
\caption{Signal Encoding(a) and Correlation Extraction Example. (b)(c)(d) Predicted as non-disease: (b) shows the voltage difference between Channel-2 and Channel-1 for Patient-1; (c) corresponds to the changes in memristor conductance associated with this voltage difference; (d) displays the correlation map.  
 (e)(f)(g) Predicted as disease: (e) shows the voltage difference between Channel-2 and Channel-1 for Patient-1; (f) corresponds to the changes in memristor conductance associated with this voltage difference; (g) displays the correlation map.}
\label{fig: tst}
\end{figure*}

\section{Evaluation}
This section details the implementation and simulation results of a seizure prediction experiment, as shown in Fig.~\ref{fig: framework}. Initially, we describe the data preprocessing steps. This is followed by a discussion on the methods used and the results obtained from the extraction of correlation features. Next, we explore the implementation and outcomes of the seizure prediction task. Finally, we present the energy consumption for the chip system and compare with other studies.
\subsection{Experimental preparation}
In epilepsy-related analysis tasks, EEG signals can be segmented into three distinct phases: \textit{ictal}, \textit{preictal}, and \textit{interictal}. The ictal samples indicate the occurrence of a seizure at the current moment. In contrast, preictal samples are used to predict the onset of an ictal phase. Interictal samples denote the data recorded between two epileptic seizures, defined as the period at least four hours before and after a seizure event\cite{truong2018convolutional}. For seizure detection tasks, a binary classification is performed between the ictal and interictal samples. Meanwhile, for seizure prediction tasks, the classification is between preictal and interictal samples\cite{li2022seizure}. Given the subtler differences between the preictal and interictal signals, seizure prediction tasks are more difficult.

\textit{\textbf{Dataset}}. We employed the CHB-MIT dataset\cite{shoeb2009application} to validate our method. This dataset comprises scalp electroencephalogram (sEEG) data from 23 pediatric patients, encompassing 844 hours of continuous sEEG recordings and 163 seizures. The sEEG signals were captured using 22 electrodes at a sampling rate of 256Hz. Following the criteria established in previous work~\cite{truong2018convolutional}, seizures occurring within 30 minutes of the last seizure were considered as a single seizure. Additionally, patient samples had to include at least two seizures and three hours of interictal period. After selection, 13 patients met our criteria, as detailed in Table~\ref{tab: patients}.

\textit{\textbf{Data Preprocessing}}. Due to the presence of invalid information in certain channels of the dataset, we selected specific channels for further analysis: ‘FP1-F3’, ‘F3-C3’, ‘C3-P3’, ‘P3-O1’, ‘FP2-F4’, ‘F4-C4’, ‘C4-P4’, ‘P4-O2’, ‘FP1-F7’, ‘F7-T7’, ‘T7-P7’, ‘P7-O1’, ‘FP2-F8’, ‘F8-T8’, ‘T8-P8’, ‘P8-O2’, ‘FZ-CZ’, ‘CZ-PZ’. These 18 channels were chosen to conduct the experiments. To mitigate the influence of noise, the raw data was filtered to a frequency range of 0-50 Hz\cite{truong2018convolutional}. Additionally, the data was segmented into 3-second windows to create samples for analysis. Finally, all memristors must be pre-set to their lowest conductance state.

\textit{\textbf{Signal Encoding}}. According to the correlation feature extraction rules described in the EEG-Extracting phase, it is necessary to encode epileptic EEG signals. The EEG signals are encoded into continuous pulse signals to preserve as much information as possible. The encoding rules, illustrated in Fig.~\ref{fig: tst} (a) , are proposed as follows: Two voltage thresholds, \( V_{pth} \) (Positive threshold) and \( V_{nth} \) (Negative threshold), are set. If the EEG signal amplitude exceeds \( V_{pth} \) or falls below \( V_{nth} \), a pulse with a voltage amplitude of 0.8V and a width of 40ns is generated. According to these rules, two types of data, Positive data and Negative data, can be obtained.

\subsection{Correlation extraction}
Following the structure depicted in Fig.~\ref{fig: framework} (b), after obtaining the Positive data and Negative data, these data sets are fed into the memristor circuit in sequence, starting with the Positive data and then the Negative data. Subsequently, correlation feature extraction is performed on the memristor chip according to the rules outlined in the EEG-Extracting stage. An example of the experimental results is shown in Fig.~\ref{fig: tst}.

Fig. \ref{fig: tst} (a) shows the Positive data represented by dark green blocks and the Negative data represented by purple blocks. Fig. \ref{fig: tst} (b) and (c) illustrate the correlation changes between Channel-2 and Channel-1 data when Patient-1 is predicted not to experience a seizure within 28.2 minutes. Fig. \ref{fig: tst} (b) depicts the voltage difference between the encoded data of Channel-2 and Channel-1 over time, with the first part representing Positive data and the second part representing Negative data. Fig. \ref{fig: tst} (c) shows the conductance values of the corresponding RRAM over time. Finally, Fig. \ref{fig: tst} (d) presents the correlation map of Patient-1 when predicted not to experience a seizure within 28.2 minutes. The conductance value at the last time point in Fig. \ref{fig: tst} (c) corresponds to the (1, 2) point in Fig. \ref{fig: tst} (d), utilizing asterisks as markers. Similarly, Fig. \ref{fig: tst} (e) and (f) describe the correlation changes between Channel-2 and Channel-1 data when Patient-1 is predicted to experience a seizure within 28.2 minutes. Fig. \ref{fig: tst} (g) presents the correlation map of Patient-1 when predicted to experience a seizure within 28.2 minutes.
\subsection{Seizure Prediction}
According to the structure depicted in Fig.~\ref{fig: framework} (c), after extracting an 18×18 correlation feature matrix on-chip, we proceed with the seizure prediction task following the rules of EEG-Computing. The prediction algorithm is an ANN consisting of only two simple layers: the first layer contains 324 weights, and the second layer contains 36 weights. This network is mapped onto the RRAM chip, as shown in Fig.~\ref{fig: chip circuit architecture} (d). 

Based on the criteria previously outlined, seizure prediction was conducted for 13 patients from the CHB-MIT dataset. The details are presented in Table \ref{tab: patients}. The average lead time for prediction was 29.2 minutes, with an average sensitivity of 91.2\%, and FPR/h of 0.11. Comparisons with other tasks are shown in Table \ref{tab: compare}, indicating that our method maintains high sensitivity and a low FPR/h in seizure prediction tasks. Additionally, there is a trade-off between FPR/h and sensitivity, which must be considered collectively during task execution. Furthermore, seizure detection and prediction are fundamentally distinct tasks. Therefore, comparisons with seizure detection are provided for reference only.

\begin{table*}[]
\centering
\footnotesize
\caption{POWER, AREA, LATENCY AND ENERGY CONSUMPTION}
\label{tab: Energy}
\renewcommand{\arraystretch}{1.2}
\setlength{\tabcolsep}{0.8mm}{
\begin{threeparttable}
\begin{tabular}{l|c|cccccc|ccccccc}
\toprule[1.2pt]
\multirow{3}{*}{\textbf{Component}} & \multirow{3}{*}{\textbf{Params.}} & \multicolumn{6}{c|}{\textbf{EEG Extracting}}                                                                                                                                       & \multicolumn{7}{c}{\textbf{EEG Computing}}                                                                                                                                                                               \\ \cline{3-15} 
                                    &                                   & \multirow{2}{*}{\textbf{Spec.}} & \textbf{Area}                    & \textbf{Power}            & \textbf{Latency}          & \textbf{T. Latency$^1$}        & \textbf{Energy}           & \multirow{2}{*}{\textbf{Spec.}} & \textbf{Area}                    & \textbf{Power}            & \textbf{Latency}          & \textbf{T. Latency}        & \textbf{Energy}           & \multirow{2}{*}{\textbf{Technology}} \\
                                    &                                   &                                 & \textbf{(mm\textsuperscript{2})} & \textbf{(mW)}             & \textbf{(us)}             & \textbf{(us)}             & \textbf{(nJ)}             &                                 & \textbf{(mm\textsuperscript{2})} & \textbf{(mW)}             & \textbf{(us)}             & \textbf{(us)}             & \textbf{(nJ)}             &                                      \\ \midrule[1pt]
WL Dirver                           & Number                            & 18                              & 2.90E-04                         & 5.50E-03                  & 2.00E-06                  & 6.14E+01                  & 3.07E-01                  & 20                              & 3.28E-04                         & 5.88E-03                  & 2.00E-06                  & 6.40E-01                  & 3.76E-03                  & CMOS(130nm)                          \\ \hline
BL Dirver                           & Number                            & 18                              & 2.90E-04                         & 5.50E-03                  & 2.00E-06                  & 6.14E+01                  & 3.07E-01                  & 36                              & 5.90E-04                         & 1.05E-02                  & 2.00E-06                  & 6.40E-01                  & 6.72E-03                  & CMOS(130nm)                          \\ \hline
SL Dirver                           & Number                            & 18                              & 2.90E-04                         & 5.50E-03                  & 2.00E-06                  & 6.14E+01                  & 3.07E-01                  & 20                              & 3.28E-04                         & 5.88E-03                  & 2.00E-06                  & 6.40E-01                  & 3.76E-03                  & CMOS(130nm)                          \\ \hline
Shift\&Adder                        & Number                            & -                               & -                                & -                         & -                         & -                         & -                         & 1                               & 2.18E-04                         & 1.00E-02                  & 1.90E-04                  & 6.40E-01                  & 6.40E-03                  & CMOS(130nm)                          \\ \hline
S\&H                                & Number                            & -                               & -                                & -                         & -                         & -                         & -                         & 40                              & 1.56E-06                         & 4.00E-04                  & 8.33E-04                  & 6.40E-01                  & 2.60E-04                  & CMOS(130nm)                          \\ \hline
MUX                                 & Number                            & -                               & -                                & -                         & -                         & -                         & -                         & 2                               & 2.72E-04                         & 7.40E-03                  & 8.00E-06                  & 6.40E-01                  & 4.74E-03                  & CMOS(130nm)                          \\ \hline
MUX decoder                         & Number                            & -                               & -                                & -                         & -                         & -                         & -                         & 2                               & 1.00E-05                         & 1.56E-04                  & 6.00E-04                  & 7.00E-01                  & 1.09E-04                  & CMOS(130nm)                          \\ \hline
\multirow{3}{*}{ADC}                & Resolution                        & \multirow{3}{*}{-}              & \multirow{3}{*}{-}               & \multirow{3}{*}{-}        & \multirow{3}{*}{-}        & \multirow{3}{*}{-}        & \multirow{3}{*}{-}        & 8 bits                          & \multirow{3}{*}{3.00E-03}        & \multirow{3}{*}{2.00E+00} & \multirow{3}{*}{1.00E-03} & \multirow{3}{*}{6.40E-01} & \multirow{3}{*}{1.28E+00} & \multirow{3}{*}{CMOS(32nm)}          \\ 
                                    & Number                            &                                 &                                  &                           &                           &                           &                           & 2                               &                                  &                           &                           &                           &                           &                                      \\ 
                                    & Sampling Speed                         &                                 &                                  &                           &                           &                           &                           & 1GS/s                           &                                  &                           &                           &                           &                           &                                      \\ \hline
ReLU                                & Number                            & -                               & -                                & -                         & -                         & -                         & -                         & 1                               & 4.80E-03                         & 1.64E-02                  & 9.80E-02                  & 9.80E-02                  & 3.22E-06                  & CMOS(22nm)                           \\ \hline
\multirow{2}{*}{eDRAM Buffer}       & Size                              & 2KB                             & \multirow{2}{*}{4.72E-03}        & \multirow{2}{*}{1.81E+01} & \multirow{2}{*}{1.15E-04} & \multirow{2}{*}{6.14E+01} & \multirow{2}{*}{1.11E+03} & 2KB                             & \multirow{2}{*}{4.72E-03}        & \multirow{2}{*}{1.81E+01} & \multirow{2}{*}{1.15E-04} & \multirow{2}{*}{6.40E-01} & \multirow{2}{*}{1.16E+01} & \multirow{2}{*}{CMOS(22nm)}          \\ 
                                    & Bus Width                         & 128                             &                                  &                           &                           &                           &                           & 128                             &                                  &                           &                           &                           &                           &                                      \\ \hline
eDRAM Bus                           & Number                            & 192                             & 4.50E-03                         & 3.50E+00                  & 9.02E-05                  & 6.14E+01                  & 2.15E+02                  & 192                             & 4.50E-03                         & 3.50E+00                  & 9.02E-05                  & 6.40E-01                  & 2.24E+00                  & CMOS(22nm)                           \\ \hline
Input Register                      & Size                              & 1KB                             & 8.10E-01                         & 6.74E-01                  & 8.21E-05                  & 6.14E+01                  & 4.14E+01                  & 1KB                             & 8.10E-01                         & 6.74E-01                  & 8.21E-05                  & 6.40E-01                  & 4.30E-01                  & CMOS(22nm)                           \\ \hline
Output Register                     & Size                              & \textbf{-}                      & \textbf{-}                       & \textbf{-}                & \textbf{-}                & \textbf{-}                & \textbf{-}                & 512B                            & 8.70E-04                         & 4.18E-01                  & 8.21E-05                  & 6.40E-01                  & 2.67E-01                  & CMOS(22nm)                           \\ \hline
\multirow{2}{*}{Crossbar}           & \multirow{2}{*}{Size}             & \multirow{2}{*}{18×18}          & \multirow{2}{*}{3.24E-06}        & \multirow{2}{*}{2.07E+00} & \multirow{2}{*}{6.00E-03} & \multirow{2}{*}{6.14E+01} & \multirow{2}{*}{1.27E+02} & 18×18                           & \multirow{2}{*}{3.60E-06}        & \multirow{2}{*}{1.44E+00} & \multirow{2}{*}{6.00E-03} & \multirow{2}{*}{3.80E-02} & \multirow{2}{*}{5.40E-02} & \multirow{2}{*}{RRAM(BEOL)}          \\
                                    &                                   &                                 &                                  &                           &                           &                           &                           & 18×2                            &                                  &                           &                           &                           &                           &                                      \\ \midrule[1pt]
\textbf{Total}                      & \textbf{-}                        & -                               & \textbf{8.20E-01}                & \textbf{2.44E+01}         & \textbf{-}                & \textbf{6.14E+01}         & \textbf{1.50E+03}         & \textbf{-}                      & \textbf{8.30E-01}                & \textbf{1.90E+01}         & \textbf{-}                & \textbf{8.36E-01}         & \textbf{1.59E+01}             & \textbf{-}                           \\ \bottomrule[1.2pt]
\end{tabular}
\begin{tablenotes}
\footnotesize
\item [1] T.Latency means total latency.
\end{tablenotes}
\end{threeparttable}

}
\vspace{-15pt}
\end{table*}

\begin{table*}[ht]
    \centering
    \subfigure{\hspace{-48pt}
    \begin{minipage}[t]{0.3\textwidth}
    \footnotesize
    \centering
    \caption{PREDICTION RESULTS}
    \label{tab: patients}
    
    
    \renewcommand{\arraystretch}{1.08}
    \setlength{\tabcolsep}{1mm}{\fontsize{8pt}{8pt}
    \begin{tabular}{cccc}
    \toprule[2pt]
    \multirow{2}{*}{\textbf{Patient}}  & \textbf{Predicted} & \textbf{Sensitivity} & \multirow{2}{*}{\textbf{FPR(/h)}} \\
                              &  \textbf{Time(min)}            & \textbf{(\%)}        &                          \\ \midrule[1pt]
    1                                 & 28.2           & 96.2       & 0.055                    \\
    2                                 & 30             & 89.1       & 0.057                    \\
    3                                 & 28.7           & 88        & 0.078                    \\
    6                                 & 28.7           & 83.3       & 0.184                    \\
    7                                & 30             & 94.2       & 0.146                    \\
    9                               & 30             & 88.6       & 0.18                     \\
    11                                & 30             & 89.2       & 0.1                      \\
    18                                & 24.2           & 97.1       & 0.202                    \\
    19                                & 30             & 86.3       & 0.084                    \\
    20                                & 30             & 96.3       & 0.042                    \\
    21                                & 30             & 89.7       & 0.16                     \\
    22                               & 30             & 94.6       & 0.062                    \\
    23                               & 30             & 92.5       & 0.08                     \\ \midrule[1pt]
    \textbf{Avg.}                           & \textbf{29.2}           & \textbf{91.2}       & \textbf{0.11}                     \\ \bottomrule[1.2pt]
    \end{tabular}}
    \end{minipage}}
    \hspace{-3pt}
    \subfigure{
    \begin{minipage}[t]{0.6\textwidth}
    \centering
    \footnotesize
    \caption{COMPARISON}
    \label{tab: compare}
    \renewcommand{\arraystretch}{0.1}
    \setlength{\tabcolsep}{0.5mm}{\fontsize{6.95pt}{6.95pt}
    \begin{tabular}{cccccccccc}
    \toprule[2pt]
    \multirow{2}{*}{\textbf{Method}} & \textbf{Number of} & \textbf{Feature}    & \textbf{Area}                    & \textbf{Latency} & \textbf{Power} & \textbf{Energy} & \textbf{Sensitivity} & \multirow{2}{*}{\textbf{FPR/h}} & \textbf{Eval.}         \\
                                     & \textbf{Patients}  & \textbf{Extraction} & \textbf{(mm\textsuperscript{2})} & \textbf{(ms)}    & \textbf{(mw)}  & \textbf{(uJ)}   & \textbf{(\%)}        &                                 & \textbf{Task(s)}       \\ \midrule[1pt]
    \multicolumn{10}{c}{\textbf{Detection}}                                                                                                                                                                                                           \\ \midrule[1pt]
    APEN,LLS{\cite{chen2013fully}}                  & 4                  & \Checkmark                   & 13.47                            & 800              & 2.8            & 77.91           & \textbf{92}          & -                               & Long Evans rats        \\
    BPF,SVM{\cite{altaf20151}}                   & 24                 & \Checkmark                   & 25                               & 2000             & 0.23           & 1.83            & 95.1                 & 0.27                            & CHB-MIT                \\
    FIR,DWT{\cite{o2018recursive}}                   & -                  & \Checkmark                   & 7.59                             & 250              & 0.674          & 168.6           & 100                  & 0.81                            & EU intracranial        \\
    FFT,SVM{\cite{huang20191}}                   & 24                 & \Checkmark                   & 4.5                              & 710              & 1.9            & 1350            & 96.6                 & 0.28                            & CHB-MIT                \\
    ICA{\cite{yang201481}}                       & -                  & \Checkmark                   & 0.4                              & 100              & 0.0816         & 8.16            & 95.24                & 0.09                            & Tzu Chi Medical Center \\
    LLS{\cite{yoo20128}}                       & 24                 & \Checkmark                   & 25                               & 2000             & 0.066          & 2.03            & 82.7                 & 0.045                           & CHB-MIT                \\ \midrule[1pt]
    \multicolumn{10}{c}{\textbf{Prediction}}                                                                                                                                                                                                               \\ \midrule[1pt]
    XGBoost{\cite{samie2018highly}}                   & 7                  & \Checkmark                   & -                                & -                & 2.42           & 12410           & 92                   & 0.039                           & Kaggle                 \\
    TDM{\cite{lammie2021towards}}                       & 5                  & \XSolidBrush                   & 0.1269                           & 1.408            & 13.3           & 187             & 78                   & 7.88                            & CHB-MIT                \\
    Parallelized{\cite{lammie2021towards}}              & 5                  & \XSolidBrush                   & 8.5089                           & 0.011            & 1700           & 187             & 78                   & 7.88                            & CHB-MIT                \\
    TDM{\cite{li2022seizure}}                       & 8                  & \XSolidBrush                   & 31.25                            & 0.445            & 2790           & 1240            & 99.24                & 0.47                            & CHB-MIT                \\
    Parallelized{\cite{li2022seizure}}              & 8                  & \XSolidBrush                   & 322.31                           & 0.0011          & 7200           & 8.12            & 99.24                & 0.47                            & CHB-MIT                \\ \midrule[1pt]
    \textbf{Ours}                    & \textbf{13}        & \textbf{\Checkmark}          & \textbf{0.83}                      & \textbf{0.0622}    & \textbf{24.4/19.01} & \textbf{1.515}   & \textbf{91.2}        & \textbf{0.11}                  & \textbf{CHB-MIT}       \\ \bottomrule[1.2pt]
    \end{tabular}}
    \end{minipage}}
    \label{fig: patients}
\end{table*}

\subsection{Energy Estimation}
The core components of our proposed chip architecture (1T1R Chip, Driver, Sample and Hold (S\&H), Shift \& Adder, Multiplexer (MUX), MUX Decoder) are based on the architecture design and parameters of a physical chip at the 130nm technology node published in\cite{yao2020fully}. The peripheral circuit components (ReLU, eDRAM Buffer, eDRAM-Tile Bus, Input Register (IR), Output Register (OR)) adopt the system framework and parameters proposed in the most relevant paper\cite{li2022seizure} at the 22nm technology node, with power calculations performed by referencing the simulation method of MemTorch \cite{lammie2022memtorch}. Specifically, the ADC component uses an 8-bit, 2 mW power consumption, and 1GS/s sampling speed high-speed low-power ADC\cite{kull20133}. The runtime for the EEG-Extracting stage is 61.44 \textmu s, and the EEG-Computing stage operates at a digital frequency of 1GHz for 836 ns. To estimate the maximum energy consumption, it is assumed that the EEG-Extracting stage continuously emits pulses (maintaining a 1.6V voltage differential on the memristor) and the EEG-Computing stage maintains a voltage of 1V (not affecting the memristor's maximum voltage). The runtime of all modules in each phase is considered to be equal to the maximum runtime.

The results are presented in Table \ref{tab: Energy}, showing a chip area of approximately 0.83 mm$^2$, a total execution time for the prediction task of approximately 62.2 \textmu s, power consumption of about 24.4 mW during the EEG-Extracting phase, and about 19.01 mW during the EEG-Computing phase. The total energy consumption for extracting correlation features from epilepsy data over a 3s window and predicting 29.2 minutes in advance is about 1.515 \textmu J. The core components are responsible for a mere 129.36 nJ of energy consumption, constituting only 8.5\% of the total, which suggests significant opportunities for optimization in the peripheral circuits, such as optoelectronic interconnect technologies. Comparisons with other results are shown in Table \ref{tab: compare}, demonstrating about 81.3\% savings in computational energy compared to the current lowest energy consumption seizure prediction tasks, and even at the estimated maximum energy limit, it remains the lowest among current seizure detection and prediction tasks. This approach can also be extended to other EEG tasks (the number of channels corresponding to array scale).

\section{CONCLUSION}
This study proposes a chip architecture and operational logic based on RRAM for reducing power consumption in correlation feature extraction and seizure prediction algorithms, enhancing the battery life of non-invasive mobile EEG collection systems. The proposed system achieves a high average sensitivity of 91.2\% and a low FPR/h of 0.11 on the CHB-MIT seizure dataset. The total energy consumption for extracting correlation features from epilepsy data over a 3s window and predicting 29.2 minutes in advance is approximately 1.515 \textmu J. The proposed method achieves a computational energy saving of about 81.3\% compared to the current lowest energy consumption (8.12 \textmu J) for epileptic seizure prediction tasks. This establishes a new benchmark for the lowest energy consumption. Due to our experiments utilizing the classic characteristics of non-volatile RRAM, such as Long-Term Potentiation (LTP) and threshold effects, this implies the possibility of further implementation on standard RRAM processes. This could facilitate the development of low-power chips designed for EEG correlation feature extraction and seizure prediction. Furthermore, the methodology can be extended to other EEG-related tasks.

\newpage
\bibliographystyle{unsrt}
\begingroup
\renewcommand{\bibfont}{\fontsize{8pt}{8pt}\selectfont} 
\bibliography{acmart}
\endgroup
\end{document}